# Thermoelastic equation of state of boron subphosphide $B_{12}P_2$

Vladimir L. Solozhenko,[a,*] Kirill A. Cherednichenko[a] and Oleksandr O. Kurakevych[b]

[a] *LSPM–CNRS, Université Paris Nord, 93430 Villetaneuse, France*
[b] *IMPMC, UPMC Sorbonne Universités, 75005 Paris, France*

Compressibility of boron subphosphide $B_{12}P_2$ has been studied under quasi-hydrostatic conditions up to 26 GPa and 2600 K using laser-heated diamond anvil cell and angle-dispersive synchrotron X-ray diffraction. 300-K data fit yields the values of bulk modulus $B_0$ = 192(11) GPa and its first pressure derivative $B_0'$ = 5.5(12). At ambient pressure the thermal expansion is quasi-linear up to 1300 K with average volume expansion coefficient $\alpha$ = 17.4(1)·$10^{-6}$ $K^{-1}$. The whole set of experimental *p-V-T* data is well described by the Anderson-Grüneisen model with $\delta_T$ = 6.

Boron subphosphide $B_{12}P_2$ is a hard (Vickers hardness $H_V$ = 35(3) GPa [1]) and refractory (melting temperature $T_m$ = 2393(30) K [2] with positive pressure slope [3]) compound with wide band gap (~2 eV [4]) and superior chemical resistance. It crystallizes in the *R-3m* space group [5], similar to α-rhombohedral boron allotrope (α-$B_{12}$) stable at high pressures [6], and other (super)hard boron-rich solids ($B_6O$, $B_{13}N_2$, $B_4C$, etc. [7-10]). Here we report the *p-V-T* equation of state (EOS) of boron subphosphide up to 26 GPa and 2600 K.

Polycrystalline powders of single-phase stoichiometric boron subphosphide were produced by self-propagating high-temperature synthesis [1] and mechanochemical synthesis [11]. The lattice parameters of synthesized $B_{12}P_2$ (*a* = 5.992(4), *c* = 11.861(8) Å) are in a good agreement with literature data (*a* = 5.9879, *c* = 11.8479 Å [5]).

At pressures 3.9-5.5 GPa and temperatures to 2000 K $B_{12}P_2$ was studied by energy-dispersive synchrotron X-ray diffraction using MAX80 multianvil system at F2.1 beamline, DORIS III (DESY). Standard assemblies with hBN pressure medium were used. The experimental details are described elsewhere [12]. Sample pressure at different temperatures was determined from thermal equation of state of hBN [13]; temperature was measured by a Pt-30%Rh/Pt-6%Rh thermocouple.

*In-situ* experiments in the 14-26 GPa pressure range have been performed in a membrane diamond anvil cell (DAC) using angle-dispersive synchrotron X-ray diffraction at P02.2 beamline, PETRA III (DESY). We used rhenium gasket and KCl pressure medium insuring quasi-hydrostatic conditions at high temperatures, with advantage of chemical inertness with regard to the sample. The monochromatic X-ray beam (42 keV, $\lambda$ = 0.2898 Å) was focused down to 2 μm×4 μm. The diffraction patterns were recorded using XRD1621 (Perkin-Elmer) flat panel detector; sample-detector distance was calibrated using $CeO_2$ NIST standard. Typical acquisition time was 30 seconds. The sample pressure has been determined using equation of state of KCl [14].

---
[*] vladimir.solozhenko@univ-paris13.fr

The 300-K $p$-$V$ data (Fig. 1a) have been fitted to Murnaghan EOS [15], i.e.

$$V(p,300) = V(0,300)\left[1 + B_0' \cdot p / B_0\right]^{-1/B_0'}, \qquad (1)$$

that allowed us to determine the isothermal bulk modulus $B_0 = 192(11)$ and its first pressure derivative $B_0' = 5.5(12)$.

Laser heating in a DAC was performed using double-sided off-axis infrared laser system (continuous fiber YAG laser focused down to 20 μm, λ = 1070 nm). Temperature measurements were performed through standard grey body radiation measurement via an Acton spectrometer SP-2356 (Princeton Instruments). The temperature uncertainties in the 1500-2500 K range were ±40 K.

High-temperature (300–1300 K) thermal expansion of $B_{12}P_2$ in argon at ambient pressure was studied at MCX beamline of Elettra synchrotron (Trieste). Debye-Scherrer geometry with rotating quartz-glass capillary was used. X-ray diffraction patterns were collected in the 5–120 2θ-range (λ = 1.0352 Å) for 120 s using a translating image plate detector upon stepwise heating with 25-K steps. Thermal expansion data (Fig. 1b) shows quasi-linear behavior following the equation $V(T)/V_0 = 1 + \alpha \cdot (T - 300)$ where $\alpha = 17.4(1) \cdot 10^{-6}$ K$^{-1}$, with a 10% higher thermal expansivity in the $c$-axis direction.

Temperature dependences of the unit-cell volume ($V_0 = V(0,300) = 366.59$ Å$^3$ corresponds to 300 K and ambient pressure) at different pressures are shown in Fig. 1b. Below 2500 K these dependences are very close to linear ones. The slopes, however, noticeably depend on pressure. To describe this dependence, we have used the thermoelastic EOS based on simplified Anderson-Grüneisen model [16] in the form

$$V(p,T) = \left[V(0,T)^{-\delta_T} + V(p,300)^{-\delta_T} - V(0,300)^{-\delta_T}\right]^{-1/\delta_T}. \qquad (2)$$

The fitted value of the Anderson-Grüneisen parameter $\delta_T = 6$ allows describing all present experimental $p$-$V$-$T$ data for $B_{12}P_2$.

Fig. 2 shows the comparison of bulk moduli of boron-rich compounds with structure related to α-rhombohedral boron. To get the correct scaling, the reported experimental data for $B_6O$ [16,17], $B_{13}N_2$ [18,19], $B_4C$ [20] and $B_{12}As_2$ [21] were fitted to Murnaghan EOS. In the case of $B_{12}As_2$ [21], we used the $p$-$V$ data up to 10 GPa only, i.e. in the range where the pressure medium used (ethanol-metanol) remains liquid, and conditions are hydrostatic. The general tendency is the decrease of bulk modulus with increase of covalent radius of an interstitial atom in the intericosahedral voids. Only boron suboxide does not follow this tendency, most probably due to the absence of boron atoms connecting oxygen atoms i.e. O-☐-O, contrary to the N-B-N and C-B-C chains in boron subnitride and carbide, respectively. *Ab initio* calculations [22,23] confirm the maximal bulk modulus for boron subnitride $B_{13}N_2$, although give overestimated $B_0$ values.

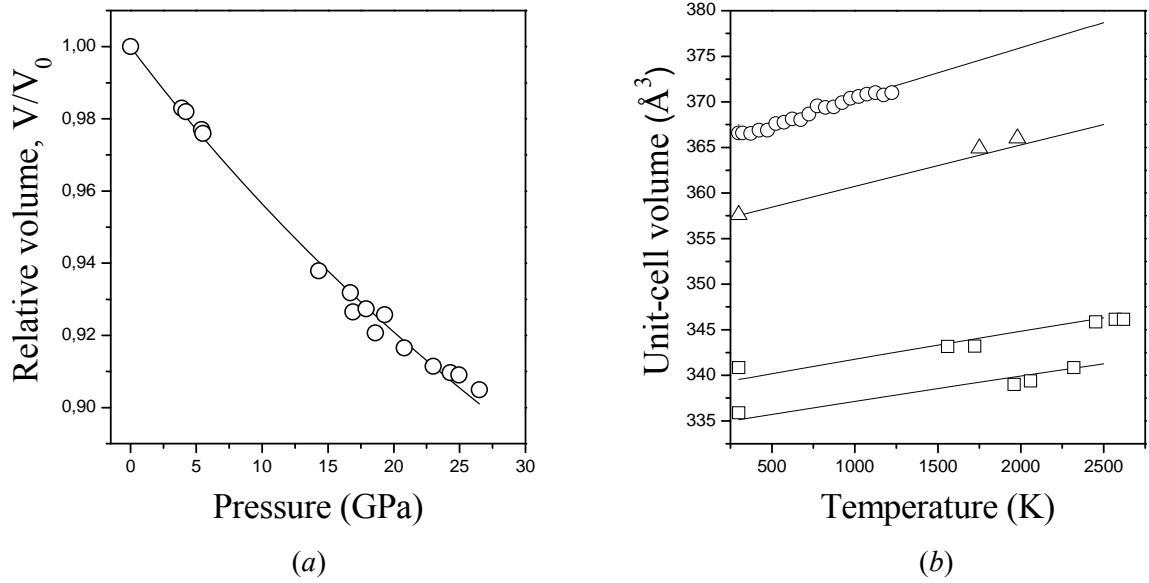

**Fig. 1** (*a*) 300-K equation of state of $B_{12}P_2$. The curve shows the data fit to Murnaghan EOS (Eq. 1). (*b*) Isobars $V(T)$ at 0.1 MPa, 5 GPa, 18 GPa and 22 GPa. The solid curves show the data fit to the simplified Anderson-Grüneisen model (Eq. 2) with $\delta_T = 6$.

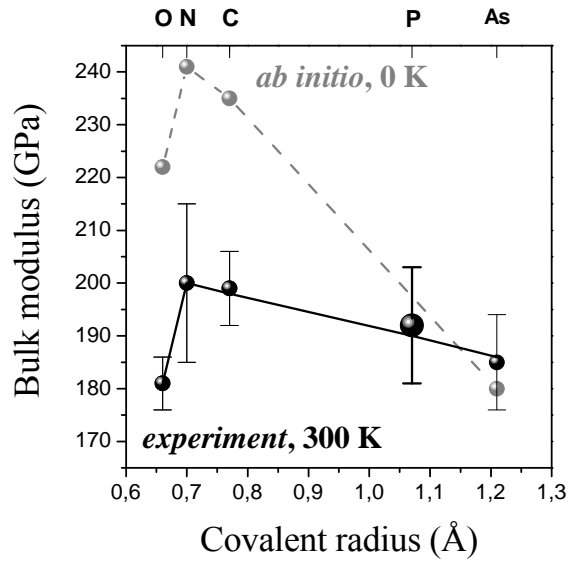

**Fig. 2** Bulk moduli of boron-rich solids with structures related to α-rhombohedral boron as a function of the covalent radius of an interstitial atom. Large ball shows the result of the present paper.

**Acknowledgements.** The authors thank Dr. V.A. Mukhanov for the samples synthesis, Dr. Y. Le Godec for the DACs preparation, and Dr. Z. Konôpková (DESY) & Dr. L. Gigli (Elettra) for assistance in the synchrotron experiments. High-pressure experiments at DESY have been carried out during beam time allocated to the Projects DESY-D-I-20090172 EC and DESY-D-I-20120021 EC and received funding from the European Community's Seventh Framework Programme (FP7/2007-2013) under grant agreement No 226716. Experiments at Elettra have been performed during beam time allocated for the Proposal No 20160086. This work was financially supported by the Agence Nationale de la Recherche (grant ANR-2011-BS08-018) and European Union's Horizon 2020 Research and Innovation Programme under Flintstone2020 project (grant agreement No 689279).